\documentclass[twocolumn,tightenlines,eqsecnum,floats,aps,amsmath,amssymb,nofootinbib,prl,shownopacs,floatfix,notitlepage]{revtex4-1}
\usepackage{graphicx, wrapfig}
\usepackage{amssymb}
\usepackage{color}
\usepackage{mathrsfs}
\usepackage{chngcntr}
\counterwithout{equation}{section}

\def\f{\frac}

\def\lp{l_{\rm Pl}}

\def\phib{\phi_{_{\rm B}}}
\def\ig{\includegraphics}

\def\planck{\emph {Planck}}

\def\lcdm{$\Lambda$CDM}
\def\klqc{k_{_{\rm (LQC)}}}
\def\lp{\ell_{\rm Pl}}

\def\aB{a(t_{\rm B})}
\def\mpc{{\rm Mpc^{-1}}}
\def\cellee{C_{\ell}^{\rm EE}}
\def\ellmax{\ell_{\rm max}}

\usepackage{enumerate}

\newcommand{\be}{\nopagebreak[3]\begin{equation}}
\newcommand{\ee}{\end{equation}}
\newcommand{\bfig}{\nopagebreak[3]\begin{figure}}
\newcommand{\efig}{\end{figure}}
\newcommand{\ba}{\nopagebreak[3]\begin{eqnarray}}
\newcommand{\ea}{\end{eqnarray}}

\newcommand{\bmult}{\nopagebreak[3]\begin{multline}}
\newcommand{\emult}{\end{multline}}
\newcommand{\fref}[1]{Fig.\,\ref{#1}}
\newcommand{\eref}[1]{eq.\,(\ref{#1})}

\begin{document}
\title{Constraining Planck scale physics with CMB and Reionization
Optical Depth}
\author{Brajesh Gupt}
\email{bgupt@gravity.psu.edu}

\affiliation{
Institute for Gravitation and the Cosmos \& Physics Department, \\The Pennsylvania
State University, University Park, PA 16802 U.S.A.
}
\affiliation{
Department of Physics and Astronomy,
Louisiana State University, Baton Rouge, LA 70803, U.S.A.
}

\begin{abstract}
We present proof of principle for a two way interplay between physics at very
early Universe and late time observations.  We find a relation between
primordial mechanisms responsible for large scale power suppression in the
primordial power spectrum and the value of reionization optical depth $\tau$. 
Such mechanisms can affect the estimation of $\tau$. We show that
using future measurements of $\tau$, one can obtain constraints on the
pre-inflationary dynamics, providing a new window on the physics of the very
early Universe. Furthermore, the new, re-estimated $\tau$ can potentially
resolve moderate discrepancy between high and low-$\ell$ measurements of $\tau$,
hence providing empirical support for the power suppression anomaly and its
primordial origin.
\end{abstract}

\maketitle

The \lcdm~model of cosmology explains up to great accuracy the temperature and
polarization spectrum of the cosmic microwave background (CMB) measured over the
past three decades. However, the recent precise measurements by the WMAP
\cite{wmap1yrsup} and \planck~\cite{planck15xvi}~missions have revealed lack of
power at large angular scales corresponding to $\ell<30$ at $\sim3\sigma$
significance level, also known as the large scale power suppression anomaly
(PSA) \cite{Schwarz:2015cma}. While its origin is still a matter of current
investigation, it is envisaged that PSA could be a relic of pre-inflationary
dynamics in the very early Universe \cite{planck15xvi}.

In this Letter, we discuss that it is possible to use the planned observational
missions to derive constraints on potential primordial mechanisms behind PSA.
Any new physics in the early Universe comes with freedom in the choice
initial conditions or physical parameters. We show that if PSA is indeed
primordial in origin, since it affects EE polarization at low-$\ell$
\cite{Das:2013sca}, it can affect the estimation of Thompson scattering optical
depth $\tau$ of late time reionization. This leads to a degeneracy between the
value of $\tau$ and the aforementioned freedom associated to primordial mechanism potentially responsible for PSA. Since the
power suppression is a low-$\ell$ phenomenon, this degeneracy can be broken via
independent measurements of the optical depth using high-$\ell$ data from future measurements. For
instance, CMB S4 mission \cite{cmbs4} and $21~{\rm cm}$ cosmology \cite{21cmtau}
corresponding to high-$\ell$ physics, are supposed to provide independent
estimations of $\tau$. Furthermore, we find that considering the suppressed
power due to primordial mechanism can alleviate a moderate discrepancy that
exists in determining mean $\tau$ from low-$\ell$ EE polarization in
\cite{planck16xlvi,planck16xlvii} and high-$\ell$ in lensed temperature data in
\cite{planck15xiii}. 

One of the most prominent estimations of $\tau$ comes from CMB via the so called
``reionization bump'' in the E-mode polarization spectrum at $\ell<20$, which
plays a crucial role in estimating $\tau$ \cite{wmap1yrtau,wmap3yrtau}. The
first constraint on $\tau$ from CMB measurements was put by the WMAP 1-year data
release to be $\tau=0.17\pm0.04$ using TE-mode polarization spectrum
\cite{wmap1yrtau} which was significantly improved by the 9-year data release to
$\tau=0.089\pm0.014$ \cite{wmap9yrtau} using the EE, TE and TT data at
low-$\ell$. In \planck~2015 data release, $\tau$ was estimated using the lensed
high-$\ell$ TT spectrum to be $\tau=0.066\pm0.016$ \cite{planck15xiii}. In
recent \planck~intermediate results, $\tau$ was re-estimated as
$\tau=0.055\pm0.009$ using the low-$\ell$ EE data coming from high frequency
instruments \cite{planck16xlvi,planck16xlvii}.  Thus, there is a moderate
discrepancy of about $\sim1.2\sigma$ between the mean value of $\tau$ from
high-$\ell$ TT and low-$\ell$ EE data by \planck. We will show that this
discrepancy can be alleviated by reestimating $\tau$ with the suppressed scalar
power spectrum.  

For explicit computations we will consider the large scale power suppression due
to the quantum gravitational corrections of loop quantum cosmology (LQC)
proposed in \cite{ag3} and show that using future measurements, we can obtain
constraints on the associated new physics in the pre-inflationary
era.\footnote{There are also other proposals for the power suppression
mechanisms
\cite{Contaldi:2003zv,Cline:2003ve,Jain:2008dw,Das:2013sca,Lello:2013mfa,Pedro:2013pba,Cai:2014vua}.
The qualitative results obtained here are expected to hold true for these
mechanisms as well.} For a given inflationary model with an inflaton field in
presence of a suitable potential in a Friedmann, Lema\^itre, Robertson, Walker
(FLRW) spacetime, LQC provides a consistent, non-singular extension of the
inflationary scenario all the way up to the Planckian curvature scale
\cite{as1,aan1,aan3}.  Let us briefly discuss the salient features of LQC
framework relevant for this paper.

\vskip0.2cm
{\bf Framework:} In the standard inflationary scenario based on classical GR,
the FLRW spacetime is described by a single spacetime metric $g_{ab}(a,\phi)$,
with $a$ being the scale factor and $\phi$ the inflaton field. We will use
the Starobinsky potential to drive inflationary dynamics (see e.g. \cite{bg2}
for a detailed analysis). However, the final results of our analysis should hold
for other choices of inflationary potential \cite{bg2,ag3}. 

In LQC, the background spacetime is given by quantum Riemannian geometry
described by a quantum wavefunction $\Psi_o(a,\phi)$ which has support on
several $g_{ab}$'s. The quantum wavefunction is obtained by solving the quantum
Hamiltonian constraint, a difference equation with the step size given by the
minimum non-zero eigenvalue of the area operator $\Delta_o$ whose value is fixed
to $\Delta_o = 5.17$ via black hole entropy computations in loop quantum
gravity. A direct consequence of the discrete quantum geometry is the resolution
of the classical big bang singularity via a quantum bounce in the expectation 
value of the scale factor \cite{aps3,as1}. The quantum bounce defines a 
characteristic LQC energy scale directly related to $\Delta_o$:
\be
\f{\klqc}{\aB} := \sqrt{\f{144 \pi^2}{\Delta_o^3}}~\lp^{-1} \approx 3.21~\lp^{-1}~,
\label{klqc}
\ee
where $\aB$ is the scale factor at the bounce and $\klqc$ is the comoving
wavenumber of the characteristic LQC scale today. While the physical wavenumber
of the characteristic LQC scale (${\klqc}/{\aB}$)at the bounce is fixed to $3.21~\lp$ in \eref{klqc}, the value
of $\klqc$ depends on specific solution to the background equations of motion
through $\aB$. There is a one parameter freedom in the choice of initial conditions that determines $\aB$ (in the convention: $a(t_0)=1$). Note that this
freedom corresponds to the freedom in the number of e-folds between the bounce
and the onset of slow-roll usually fixed by choosing the value of inflaton at
the bounce $\phib$. As discussed in \cite{ag3}, physical principles rooted in
the simplest quantum geometry can be used to fixed this freedom as $\aB\approx e^{-141}$. That, in turn, yields:
\be
\klqc = 0.00024~\mpc~.
\label{klqc2}
\ee
Note that \eref{klqc2} represents the value expected from the simplest quantum
geometry in the deep quantum gravity. If this assumptions is dropped, $\klqc$
becomes a free parameter and would need to be refined using inputs from
observations.

\vskip0.2cm
{\bf CMB Polarization spectrum:} $\klqc$ defines a scale at which the
pre-inflationary effects to the power spectrum become important. Modes of
cosmological perturbations with $k\gg\klqc$ remain unaffected by LQC 
corrections. However, the infrared modes with $k\lesssim \klqc$ carry an imprint
of the quantum gravity era and arrive at the slow-roll phase in an excited state
\cite{aan3}. As shown in \cite{ag3}, with the appropriate choice of initial
conditions for perturbations, the power spectrum of these modes is significantly
different from the standard one and is suppressed at scales corresponding to
multipoles $\ell <30$. The resulting temperature-temperature spectrum then fits
better with the Planck data than the one corresponding to the standard, nearly 
scale invariant primordial power spectrum (PPS).

\bfig
\ig[width=0.5\textwidth]{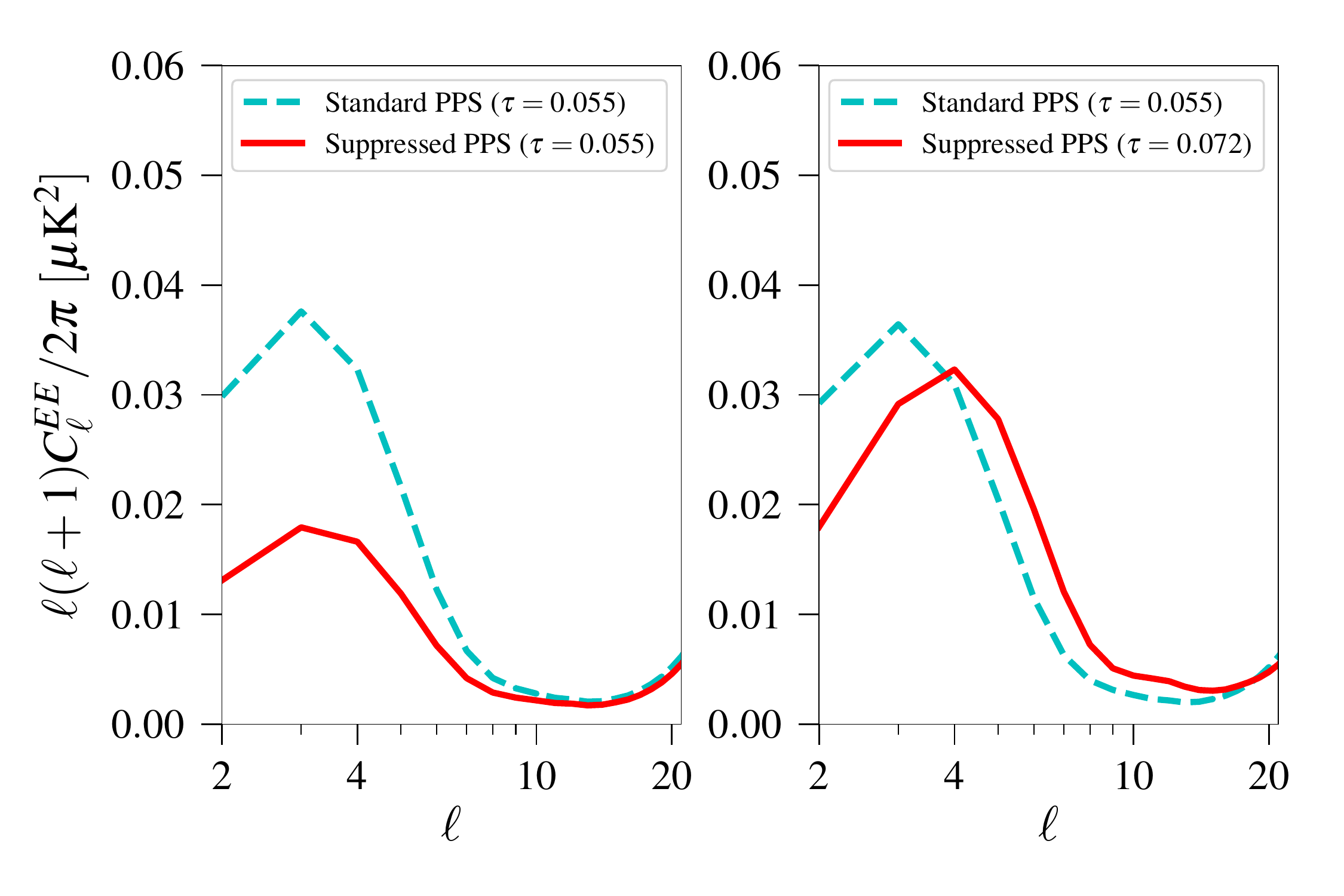}
\caption{E-mode polarization spectrum for $\ell=2-20$ for the standard (dashed blue)
and suppressed LQC primordial power spectra (solid red). In the left panel both
curves correspond to $\tau=0.055$, while in the right panel $\tau$ has been increased to
$0.072$ for the suppressed LQC PPS.}
\label{cellEE}
\efig
For the analysis in this paper, we will restrict ourselves to the EE spectrum
for $\ell=2-20$, similarly to the recent analysis of the reionization history by
\planck~ \cite{planck16xlvi,planck16xlvii}. As discussed in
\cite{planck16xlvii}, this is enough as the high-$\ell$ likelihoods in EE do not
contain additional information about reionization. \fref{cellEE} shows the EE
polarization spectrum for the standard PPS and the suppressed PPS of LQC, where
the LQC characteristic scale is fixed as in \eref{klqc2}. The left panel
compares the power spectra for $\tau=0.055$, the best fit value obtained in
\cite{planck16xlvi}, while all the other cosmological parameters are fixed to
their best fit values reported in \cite{planck15xiii}. The amplitude of the
reionization bump for the LQC spectrum is suppressed. The right panel compares
the polarization spectra for standard PPS with $\tau=0.055$ to the suppressed
LQC PPS with $\tau=0.072$. This implies that, {\it the
suppressed power spectrum would predict a larger value of the optical depth}.
Thus, there is an apparent correlation between the values of $\klqc$ and $\tau$.

\vskip0.2cm
{\bf Fisher information matrix and error bars:} 
\bfig
\ig[width=0.5\textwidth]{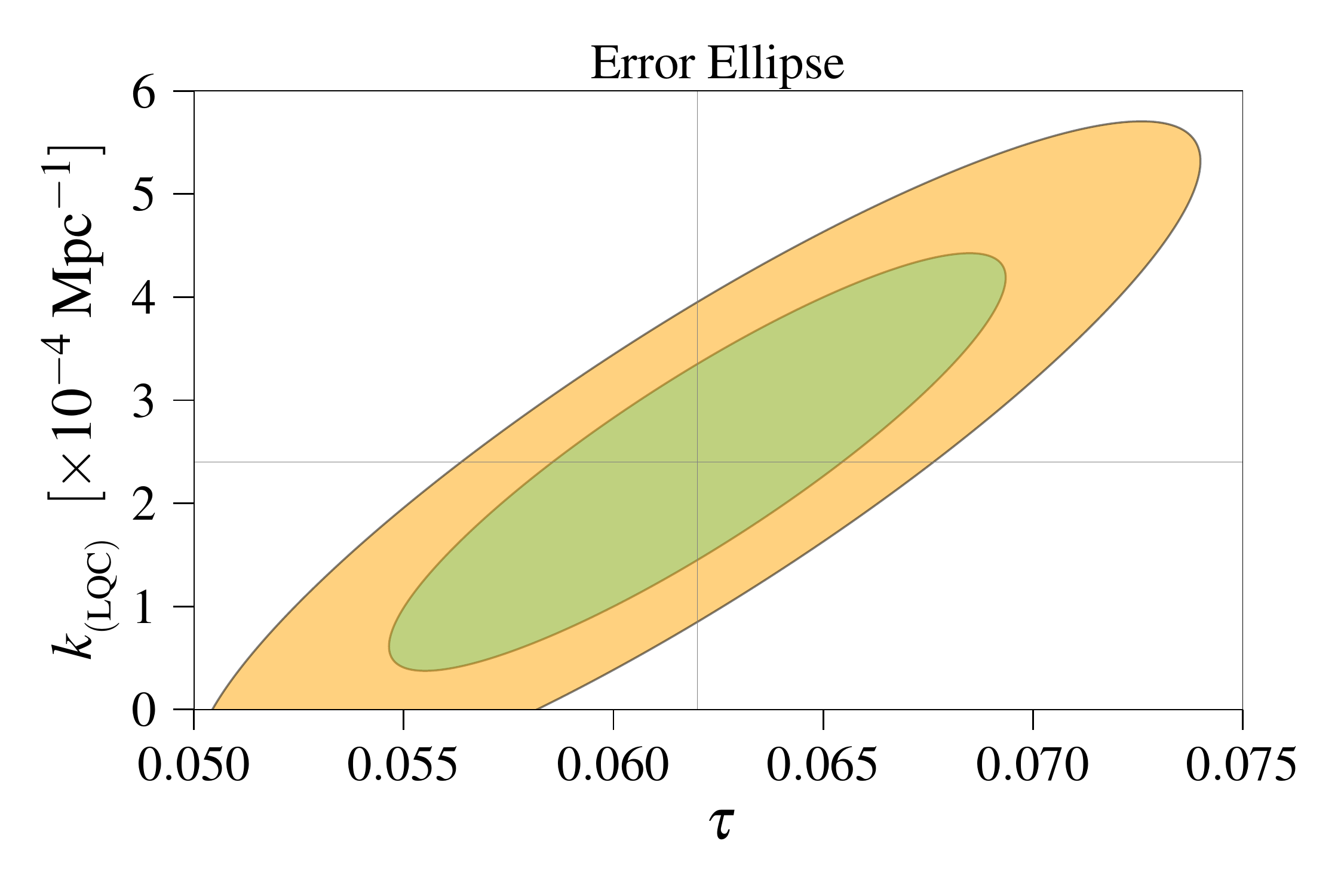}
\caption{Error ellipse for $\klqc$ and $\tau$. The inner and outer ellipses show 68
and 95\% error contours respectively. The mean value is taken to be
$\klqc=0.00024~\mpc$ and the corresponding value of optical depth $\tau=0.062$ determined 
using the suppressed LQC PPS.}
\label{fisherfig}
\efig
To quantify the aforementioned correlation $\klqc$ and $\tau$, we compute the 
Fisher information matrix \cite{fisher}. Since the \planck~ data for low-$\ell$
polarization is not yet available, for this analysis we will assume that the
error bars on $\cellee$ at low-$\ell$'s is given by the cosmic variance limit.
As discussed in \cite{tegmarkfisher}, the Fisher matrix then takes the following
form:
\be
 \mathbb F_{ij} = \sum_{\ell=2}^{\ellmax} \f{1}{\left(\Delta \cellee\right)^{2}} 
                         ~\f{\partial \cellee}{\partial \theta_i}
                         \f{\partial \cellee}{\partial \theta_j},
\label{fishereq}
\ee
where 
\be
 \Delta \cellee = \sqrt{\f{2}{2\ell+1}}~\cellee,
\ee
and $\theta = \left(\tau,~\klqc\right)$, while the other cosmological parameters are
fixed at their best fit value given in \cite{planck15xiii}. Recall that effects of
$\klqc$ are limited to very large angular scales and $\tau$ is determined from the
E-mode reionization bump which also occurs at $\ell<20$. Therefore, we keep $\ellmax$ in
\eref{fishereq} large enough (at least $50$) to include the affected multipoles 
in our analysis.

Elements of the covariance matrix between $\klqc$ and $\tau$ are then obtained by
inverting the Fisher matrix: $\mathbb C_{ij}=\left(\mathbb F^{-1}\right)_{\ij}$.
\fref{fisherfig} shows the error ellipse corresponding to $\mathbb C_{ij}$. The inner
and outer contours correspond to the $68\%$ and $95\%$ confidence levels respectively. As
expected from previous discussions, there is a strong degeneracy between $\klqc$ and
$\tau$. Note that $\mathbb C_{ij}$ only captures the information about the error bars and
correlation between the two parameters. The mean values of $\tau$ and $\klqc$ at which 
the errors ellipse is centered are given by the best fit values which
we have obtained by proceeding as follows.

\vskip0.2cm
{\bf Implications for future observations and Constraints on LQC:}  
As evident from \fref{fisherfig}, there is strong degeneracy between $\klqc$ and
$\tau$ measured from the low-$\ell$ polarization data. In order to break this
degeneracy we would need an independent estimation of either $\klqc$ or $\tau$.
The LQC scale $\klqc$ is a parameter of the underlying theory. On the other hand, 
$\tau$ can be measured using high-$\ell$ TT data as well as by upcoming 
experiments such as CMB S4 \cite{cmbs4} and 21 cm cosmology \cite{21cmtau} missions 
independent from \planck~measurements. {\it The measured value of $\tau$ from these
experiments will break the degeneracy with $\klqc$ and put observational
constraints on $\klqc$}. 

Recall that the value of $\klqc$ determines $\aB$, i.e. the
initial conditions of the background geometry at the bounce.\footnote{This assumes
that the state for quantum perturbations are fixed using a quantum
generalization of the Weyl curvature hypothesis as discussed in \cite{ag2,ag3}}
Thus, we can learn about the properties of quantum geometry using future observational
data. Moreover, as discussed before, since the suppressed power determines a higher value
of $\tau$, it can resolve the moderate discrepancy in the estimation of $\tau$ using the
low-$\ell$ polarization and high-$\ell$ temperature data from \planck~mission.

\vskip0.2cm
{\bf Re-estimating Optical Depth:}
\bfig
\ig[width=0.5\textwidth]{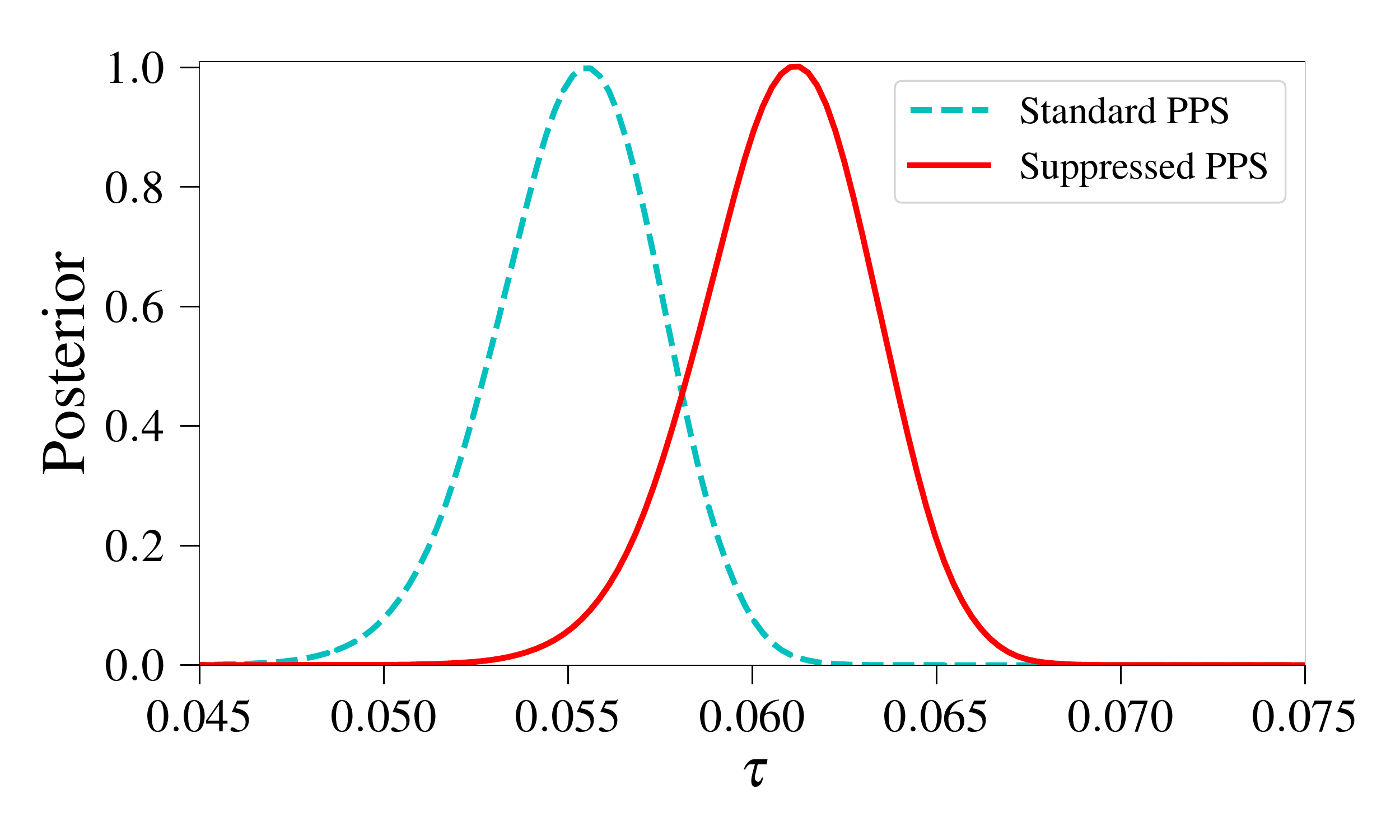}
\caption{1-dimensional posteriors of $\tau$ obtained with the ``simulated'' data for 
the standard PPS (dashed blue curve) and the suppressed LQC PPS (solid red curve)
with $\klqc=0.00024~\mpc$, which is chosen to represent the mean of the error ellipse
in \fref{fisherfig}. }
\label{1dlike}
\efig
Note that, using the low-$\ell$ TT data in \cite{planck16xlvi}, $\tau$ was
determined to be $0.055$. Therefore, it is possible that if the suppressed power
spectrum is used to determine $\tau$ with the low-$\ell$ EE data, the new value
of $\tau$ might increase enough and come closer to $0.066$---the value obtained
from high-$\ell$ TT data \cite{planck15xiii}---hence resolving an apparent
discrepancy between estimation of $\tau$ from low-$\ell$ EE and high-$\ell$ TT
data. Let us find out if this expectation is borne out in our analysis.

To obtain the best fit values of $\tau$ for the suppressed PPS we perform the
maximum likelihood analysis with low-$\ell$ $\cellee$ by varying $\tau$, keeping
$\klqc=0.00024~\mpc$ (\eref{klqc2}), obtained from consideration of simplest quantum geometry
in the deep Planck regime \cite{ag3}, while fixing other cosmological parameters
at their best fit values given in \cite{planck15xiii}.

In order to perform this analysis, we need the EE spectrum measured from experiments
at low-$\ell$ which, however, has not been made available publicly
yet. Given the lack of real data, we will work with a ``simulated''
EE data at low-$\ell$ constructed in the following manner. We fix $\tau=0.055$ 
(i.e.\ the mean value of $\tau$ obtained in \cite{planck16xlvi}) and compute
$\cellee$ assuming the standard almost scale invariant power spectrum. We consider
this to be the central values of $\cellee$ with the errors bars given by the cosmic
variance at low-$\ell$.\footnote{Cosmic variance is the theoretical lower limit on
the error bars at
low-$\ell$ and no observational can beat it. However, see \cite{Kamionkowski:1997na}
for potential way of getting around cosmic variance via careful measurements of
quadrapole of the galaxy clusters and CMB secondaries.} In this sense, our
``simulated'' data represents the best ever possible CMB measurements at low
multipoles.\footnote{Of
course, in the real data due to instrumental noise and systematics the real error
bars would be larger which we will revisit when the low-$\ell$ polarization data from
\planck~is released. We performed an estimation of the effect of higher error bars on 
the degeneracy found here by digitizing figure 33 of \cite{planck16xlvi}. We found that 
the degeneracy is not affected by larger error bars.}

\fref{1dlike} shows the corresponding one dimensional posterior distribution of $\tau$ 
for the standard PPS (dashed blue curve) and the suppressed LQC PPS (solid red
curve).  The estimated $\tau$ with $95\%$ error bars are:
\ba
  \tau &=& 0.055\pm 0.008\qquad\qquad {\rm (Standard~PPS) } \nonumber \\
  \tau &=& 0.062\pm 0.008\qquad\qquad {\rm (Suppressed~PPS) }.
\ea
Note that the width of the posterior distribution is sharper as compared
to that obtained in \cite{planck16xlvi}, because here we have considered the
``simulated'' data with cosmic variance error bars which are significantly smaller.
It is evident that the peak of the posterior has shifted to a higher value of $\tau$
when the suppressed power spectrum is used. Moreover, the re-estimated value is closer to
the one obtained from high-$\ell$ TT data: $\tau=0.066$. This indicates that if 
the large scale power suppression is indeed a primordial effect rather than a statistical
fluke, it can resolve the discrepancy between the estimation of $\tau$ purely from
low-$\ell$ EE polarization spectrum (estimated to be $0.055$ with the standard PPS) and 
high-$\ell$ TT spectrum. {\it This indicates further empirical support for the 
possibility that the PSA could have originated from physical processes in the very 
early Universe.}

\vskip0.5cm

In this Letter we have shown that future observational data, in particualr
giving independent measurement of $\tau$ can be used to determine the scale at
which PSA is observed in the TT spectrum, which in turn can constrain the
associated pre-inflationary physics. Here we have only presented a proof of 
principle that there is potentially
a new window on pre-inflationary physics via a symbiotic interplay between
observational data and fundamental. While we performed a case study by
considering the pre-inflationary dynamics of loop quantum cosmology, the overall
results of the analysis can be extended to other primordial mechanisms which
introduces a characteristic scale for suppression of power at large angular
scales. Due to the lack of availability of
observational data for polarization at low-$\ell$ we used simulated data
assuming the error bars on $\cellee$ at low-$\ell$ to be given by the cosmic
variance. The actual experimental data from \planck~expected to be released in
upcoming few months, will have higher error bars which is expected to only
increase the width of the error ellipse while keeping the degeneracy intact. We
will revisit this analysis when more data from \planck, CMB S4 and 21 cm
cosmology is available, which hopefully will provide new observational insights
on the physics of deep Planck regime in the very early Universe.

\vskip0.2cm
\acknowledgements
{\bf Acknowledgements:} We would like to thank Nishant Agarwal, Abhay Ashtekar, 
Donghui Jeong and Suvodip 
Mukherjee for comments, suggestions and disucssions and Ivan Agullo, B\'eatrice Bonga, Anne Sylvie Deutsch, 
Anuradha Gupta, Charles Lawrence, and Tarun Souradeep for helpful discussions. This 
work was supported by NSF grant PHY-1505411 and the Eberly research funds of Penn
State, and in part by Grant No. NSF-PHY-1603630, funds of the
Hearne Institute for Theoretical Physics and CCT-LSU.
This work used the Extreme Science and Engineering Discovery Environment (XSEDE), which 
is supported by National Science Foundation grant number ACI-1053575. This work was
supported 

\section*{References}

\end{document}